\documentclass[aps,pra,twocolumn,superscriptaddress,reprint,amsmath,amssymb,floatfix,eqsecnum]{revtex4}
\usepackage{graphicx}
\usepackage{amssymb,bbm,amsmath,braket,indentfirst,bm}
\allowdisplaybreaks 
\usepackage[pdftex]{color} 
\usepackage{hyperref}
\usepackage{color}
\usepackage{ulem} 
\usepackage[paperwidth=8.5in,paperheight=11in,centering,hmargin=2cm,vmargin=2.5cm]{geometry} 

\newcommand{\pri}[1]{\ensuremath #1^{\prime}}

\renewcommand\({\ensuremath \left(} \renewcommand\){\ensuremath
  \right)} \renewcommand\[{\ensuremath \left[}
  \renewcommand\]{\ensuremath \right]}
\def\:={\,\raisebox{0.85pt}{.}\hspace{-2.78pt}\raisebox{2.85pt}{.}\!\!=\,}
\def\=:{\,=\!\!\raisebox{0.85pt}{.}\hspace{-2.78pt}\raisebox{2.85pt}{.}\,}

\newcommand{\ti}{\tilde{i}}
\newcommand{\tj}{\tilde{j}}
\newcommand{\tk}{\tilde{k}}
\newcommand{\tl}{\tilde{l}}

\begin{document}

\title{Occupation of topological Floquet bands in open systems}

\author{Thomas~Iadecola} \affiliation{Physics Department, Boston
  University, Boston, Massachusetts 02215, USA}

\author{Titus~Neupert} \affiliation{Princeton Center for Theoretical
  Science, Princeton University, Princeton, New Jersey 08544, USA}

\author{Claudio~Chamon} \affiliation{Physics Department, Boston
  University, Boston, Massachusetts 02215, USA}

\date{\today}

\begin{abstract}
Floquet topological insulators are noninteracting quantum systems that, when driven by a time-periodic field, are described by effective Hamiltonians whose bands carry nontrivial topological invariants.  A longstanding question concerns the possibility of selectively populating one of these effective bands, thereby maximizing the system's resemblance to a static topological insulator.  We study such Floquet systems coupled to a zero-temperature thermal reservoir that provides dissipation.  We find that the resulting electronic steady states are generically characterized by a finite density of excitations above the effective ground state, even when the driving has a small amplitude and/or large frequency.  We discuss the role of reservoir engineering in mitigating this problem.
\end{abstract}

\maketitle

\section{Introduction}

In recent years, the possibility of engineering topological states of matter in otherwise trivial materials has motivated significant interest in electronic systems driven periodically in time~\cite{oka,lindner,kitagawa,fertig,rudner,grushin,kundu,torres,moore}.  The prescription for a particular target topological state is obtained using Floquet theory~\cite{shirley,sambe}, which describes a periodically-driven quantum system with Hamiltonian $H(t)=H(t+\tau)$ in terms of a time-independent Hamiltonian $H_{\rm F}$.  In particular, Floquet's theorem states that there exists a complete basis of solutions of the time-dependent Schr\"odinger equation $H(t)\ket{\psi_j(t)}=\mathrm i\, \partial_t\ket{\psi_j(t)}$ of the form
\begin{align}\label{floquet states}
\ket{\psi_j(t)}=e^{-\mathrm i\epsilon_j t}\, \ket{u_j(t)},
\end{align}
where the state $\ket{u_j(t)}=\ket{u_j(t+\tau)}$ is periodic in time with the same period as $H(t)$.  The quasi-energies $\epsilon_j$ can be thought of as the energy spectrum of an effective ``Floquet Hamiltonian" $H_{\rm F}$ defined by evaluating the evolution operator $U(t_2,t_1)$ at an integer multiple of the period $\tau$, i.e.~$U(n\tau,0)=:\exp(-\mathrm i\, H_{\rm F}\, n\tau)$.  Floquet's theorem guarantees the existence of a time-periodic unitary operator $P(t)$ that maps the time-dependent Hamiltonian onto $H_{\rm F}$ \cite{rahav,goldman,bukov}:
\begin{align}\label{rotating frame}
H_{\rm F}=P^\dagger(t)H(t)P(t)-\mathrm i\, P^\dagger(t)\partial_t P(t).
\end{align}
If the single-particle Hamiltonian $H_{\rm F}$ and its associated eigenstates $\ket{u_j(n\tau)}$ define a model with nontrivial Berry curvature, then it is possible for the quasienergy ``bands" to possess nonzero Chern numbers or $\mathbb Z_2$ invariants~\cite{oka,lindner}, defined by analogy with static Hamiltonians~\cite{hasan}.

\begin{figure}
\centering
\includegraphics[width=.38\textwidth,page=2]{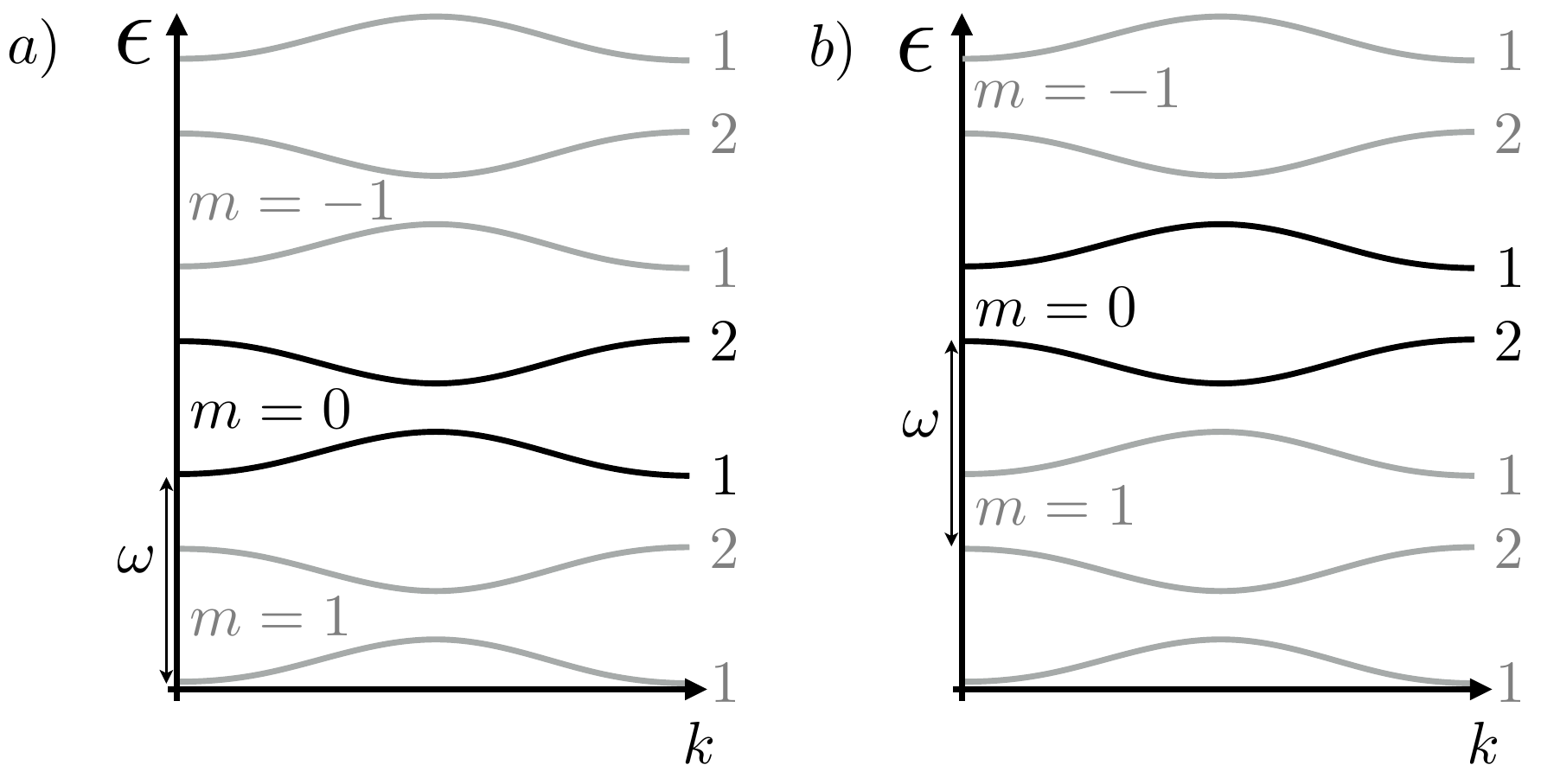}
\caption{Ambiguity in the ordering of quasienergy bands.  For two quasienergy bands falling within a given strip of size $\omega$, as in a), a gauge transformation of the form \eqref{gauge} can be used to shift the window and invert the ordering, as in b).   }\label{fig:gauge}
\end{figure}

Despite the formal similarity to undriven systems, there are several fundamental differences between a Floquet effective Hamiltonian $H_{\rm F}$ and a static Hamiltonian.  First, there is an ambiguity in the definition of $H_{\rm F}$ above: the quasienergies $\epsilon_j$ are only defined modulo multiples of the driving frequency $\omega=2\pi/\tau$.  In other words, the definition of $H_{\rm F}$ is invariant under the gauge transformation
\begin{align}\label{gauge}
\epsilon_j\to \epsilon_j+n_j\, \omega\indent \text{and}\indent \ket{u_j(t)}\to e^{\mathrm i n_j\omega t}\ket{u_j(t)},
\end{align}
for any set of integers $\{n_j\}$.  This makes an {\it a priori} definition of a ``lowest" quasienergy impossible, since one can always fold and/or reorder (c.f.~Fig.~\ref{fig:gauge}) the spectrum by means of such a gauge transformation~\cite{kohn}.

This obstruction to defining a unique Floquet ground state highlights a second important contrast with static systems, namely the fact that there is no universal principle determining the occupations of quasienergy states at zero temperature.  Indeed, this signature of the inherently out-of-equilibrium nature of periodically-driven systems poses a challenge to theoretical proposals of Floquet topological states --- even if a Floquet system has topological quasienergy bands, there is no guarantee that the long-time state of the system is a pure Floquet state with the desired features~\cite{moore, dehghani1, dehghani2}.  For open systems in contact with a thermal reservoir, as is the case in solid-state realizations, the detailed properties of the reservoir and its coupling to the system play crucial roles in determining the nonequilibrium steady state of the system at long times~\cite{grifoni,kohler,breuer,kohler_review,kohn,hone,grand_floquet,seetharam}.

In this work, we investigate the occupations of topological Floquet bands in a noninteracting fermionic system coupled to a thermal reservoir at zero temperature.  We ask whether, and under what conditions, it is possible for the nonequilibrium steady state of the system to feature a single fully populated Floquet band when the original undriven system is half-filled.  Using a Floquet master equation approach, we examine as a function of the driving amplitude and frequency the rate to escape a given Floquet state. We find that generic systems undergo heating that spoils the complete occupation of a single band, even in the favorable limit of weak, off-resonant driving.  In the best-case scenario, the excitation density falls off as a power law in $1/\omega$, with the non-universal decay exponent set by the low-energy behavior of the bath density of states.  We close by suggesting bath engineering schemes that could further suppress excitations.  While our motivation derives from the pursuit of topological phases in driven systems, we point out that our results hold equally well in ``trivial" Floquet band insulators.

The structure of the paper is as follows.  In Sec.~\ref{master equation review}, we provide an overview of the Floquet master equation approach used in this work.  Rather than focusing on exposition, this section provides a fresh perspective on the subject by emphasizing the requirement of invariance of all physical quantities under gauge transformations of the form \eqref{gauge}.  This physical principle resolves many of the conceptual ambiguities regarding the definitions of Floquet states and the associated quasienergies.  Within this formalism, we outline conditions under which open Floquet systems can reach steady states that resemble their equilibrium counterparts, i.e.~where the occupations of the Floquet states at finite temperature are distributed in a Boltzmann-like fashion.  In the remaining sections, which can be read independently of Sec.~\ref{master equation review}, we specialize to the study of noninteracting, two-band Floquet systems coupled to zero-temperature bosonic reservoirs, and present the main results of this work, which were outlined above.

\section{Floquet master equations}\label{master equation review}

\subsection{Definitions and comments on gauge invariance}\label{gauge invariance review}
In this paper, we will study Floquet systems that are described by Hamiltonians of the form
\begin{align}\label{model def}
H(t) = H_{\rm S}(t)+H_{\rm SB}+H_{\rm B},
\end{align}
where $H_{\rm S}(t)=H_{\rm S}(t+\tau)$ describes the periodically driven system, $H_{\rm B}$ describes the reservoir (or ``bath"), and $H_{\rm SB}$ describes the coupling between them.  Following Refs.~\onlinecite{kohn} and \onlinecite{hone}, we assume a system-bath coupling of the factorized form
\begin{align}\label{H_SB def}
H_{\rm SB} = \gamma\, S\, B, 
\end{align}
where $\gamma$ is a real coupling constant [assumed to be smaller than any energy scale in $H_{\rm S}(t)$], and where $S$ and $B$ are Hermitian operators acting solely on the degrees of freedom of the system and the bath, respectively~\cite{footnote_sys-bath}.  System-bath couplings of this form are ubiquitous in the study of open quantum systems, e.g.~in spin-boson-type models~\cite{leggett}.  The reservoir described by $H_{\rm B}$, whose eigenstates and energies $H_{\rm B}\ket{\nu}=E_\nu\ket{\nu}$ are known, is assumed to be in thermal equilibrium (either at zero temperature or at a finite temperature $1/\beta$, although in Sec.~\ref{nonthermal} we will specialize to the case of zero temperature).

The unitary time evolution of the full \textit{closed} system is completely characterized by the density matrix $\varrho(t)$, whose equation of motion is given by
\begin{align}\label{von Neumann equation}
\mathrm i \partial_t\, \varrho(t) = [H(t),\varrho(t)].
\end{align}
Our interest, however, is primarily in the influence of the reservoir on the system described by $H_{\rm S}(t)$.  The important quantity to consider is then the reduced density matrix
\begin{align}\label{reduced density matrix}
\rho(t) := \text{Tr}_{\rm B}\[\varrho(t)\],
\end{align}
where $\text{Tr}_{\rm B}[\ \cdot\ ]$ represents the operation of taking the trace over the bath degrees of freedom.  The evolution of the \textit{open} system described by $\rho(t)$ is obtained by ``tracing out" the bath in Eq.~\eqref{von Neumann equation}, which results in a master equation for the reduced density matrix.  Owing to the explicit time-dependence of $H_{\rm S}$, this master equation has time-periodic coefficients, so that the Floquet states defined by Eq.~\eqref{floquet states} form a natural basis in which to resolve $\rho(t)$~\cite{grifoni,kohler,breuer,kohler_review,kohn,hone}.  The derivation of the master equation for the reduced density matrix in the Floquet basis proceeds via the Born-Markov approximation, which amounts to an assumption of weak coupling between the system and bath, as well as the assumption that the correlation time of the bath is sufficiently short that memory effects can be neglected.   

A particularly clear derivation of this Floquet master equation, as well as a careful discussion of the hierarchy of time-scales necessary in order for the Born-Markov approximation to hold, is presented by Hone et al.~in Ref.~\onlinecite{hone}.  We refer the reader to that work (and references therein) for details, and begin our discussion with the master equation written in their notation:
\begin{subequations}
\begin{align}\label{gauge inv master equation}
\(\partial_t+\mathrm i \epsilon_{ij}\)\rho_{ij}(t) &=  -\frac{1}{2}\sum_{k,l}\left\{\rho_{lj}(t)R_{ik;lk}(t)+\rho_{il}(t)R^*_{jk;lk}(t)\right.\nonumber\\
&\indent\left.-\rho_{kl}(t)\[R_{lj;ki}(t)+R^*_{ki;lj}(t)\]\right\},
\end{align}
where 
\begin{align}
\rho_{ij}(t)&:=\bra{u_i(t)}\rho(t)\ket{u_j(t)},\\
\epsilon_{ij}&:=\epsilon_i-\epsilon_j,
\end{align} and the coefficients $R_{ij;kl}(t):= \sum_{K}e^{iK\omega t}\, R_{ij;kl}(K)$, where
\begin{align}
R_{ij;kl}(K)&:= 2\pi\gamma^2\sum_{m} S_{ij}(m+K)\, S^*_{kl}(m)\, g(\epsilon_{lk}-m\omega)\nonumber\\
S_{ij}(m) &:= \frac{1}{\tau}\int_0^\tau\mathrm d t\, e^{-\mathrm i m\omega t}\, \bra{u_i(t)}S\ket{u_j(t)}.
\end{align}
The influence of the bath is contained in the function $g(E)$, essentially a weighted density of states, which is defined by
\begin{align}\label{g(E) definition finite T}
g(E)&=\frac{1}{Z}\sum_{\mu,\nu} e^{-\beta E_\mu}|\langle \mu|B|\nu\rangle|^2 \,\delta(E+E_{\nu}-E_\mu),\\
Z &= \sum_\nu e^{-\beta E_\nu},
\end{align}
\end{subequations}
where $\beta$ is the inverse temperature.
 
Equation~\eqref{gauge inv master equation} is advantageous in that it is completely invariant under the gauge transformations \eqref{gauge} that reshuffle the quasienergy spectrum.  Indeed, defining
\begin{subequations}\label{transformed states and quasienergies}
\begin{align}
\ket{\pri u_j(t)}&:= e^{\mathrm in_j\omega t}\ket{u_j(t)}\\
\pri\epsilon_j&:=\epsilon_j+n_j\omega,
\end{align}
\end{subequations}
one finds that the density matrix transforms as
\begin{align}\label{density matrix transformation}
\rho_{ij}(t) =e^{\mathrm i n_{ij}\omega t}\, \rho^\prime_{ij}(t)\equiv e^{\mathrm i (n_i-n_j)\omega t}\, \bra{u^\prime_i(t)}\rho(t)\ket{u^\prime_j(t)},
\end{align}
while the time-dependent rates transform as
\begin{align}
R_{ij;kl}(t)&=e^{\mathrm i(n_{ik}-n_{jl})\omega t}\, R^\prime_{ij;kl}(t).
\end{align}
Consequently, the left- and right-hand sides of Eq.~\eqref{gauge inv master equation} transform with an oscillating phase factor that cancels, and the equation is therefore invariant under the gauge transformation.

Despite the appeal of a gauge-invariant equation of motion for the reduced density matrix, it is also difficult to make analytical progress while the rates $R_{ij;kl}(t)$ are time-dependent.  If one is only interested, as we are, in the long-time limit, where the density operator does not vary substiantially over a single period, then a valid means of bypassing this difficulty is to consider the {\it time-average} of both sides of Eq.~\eqref{gauge inv master equation}~\cite{hone},
\begin{subequations}
\begin{align}\label{averaged master equation}
\begin{split}
\mathrm i\, \epsilon_{ij}\, \rho_{ij} &= -\frac{1}{2}\sum_{k,l}\[\rho_{lj}R_{ik;lk}+\rho_{il}R^*_{jk;lk}\right.\\
&\qquad\qquad\qquad \left.-\rho_{kl}\(R_{lj;ki}+R^*_{ki;lj}\)\],
\end{split}
\end{align}
where $\rho_{ij}\equiv \overline{\rho_{ij}(t\to\infty)}$ is the time-average of $\rho_{ij}(t)$, and where
\begin{align}
R_{ij;kl}=R_{ij;kl}(K=0)
\end{align}
\end{subequations}
are the time-averaged rates.  Equation~\eqref{averaged master equation} can be further simplified by considering the structure of the averaged density matrix $\rho_{ij}$.  Indeed, so long as $\epsilon_{ij}=\epsilon_i-\epsilon_j\neq 0$ for $i\neq j$, the system-bath coupling $\gamma$ can be chosen to be much smaller than the smallest $\epsilon_{ij}$.  In this case, the off-diagonal elements of the steady-state reduced density matrix vanish to order $\gamma^2$ (c.f.~\cite{hone}), and the occupations $p_i:=\rho_{ii}$ of the Floquet states are solutions to the rate equation
\begin{align}\label{full rate equation}
0=\sum_{j\neq i}\(R_{j\to i}\, p_j-R_{i\to j}\, p_i\),
\end{align}
where the transition rates
\begin{eqnarray}
R_{i\to j}
&=&
R_{i j;i j}
=
R^*_{i j;i j}\nonumber
\\
&=&
2\pi\gamma^2
\sum_m \left|S_{i j}(m)\right|^2\,g(\epsilon_{ji}-m\omega).
\end{eqnarray}
Our treatment of nonthermal steady states in Sec.~\ref{nonthermal} will be based on a careful analysis of the rate equation \eqref{full rate equation} in the special case of a two-state system.

The time-averaging process outlined above spoils the gauge-invariance of Eq.~\eqref{gauge inv master equation};  after time-averaging, one therefore implicitly commits to a choice of ordering for the quasienergies.  Indeed, observe that
\begin{align}
\begin{split}
R_{ij;kl}(K)
&=R^\prime_{ij;kl}(K+n_{ik}-n_{jl}).
\end{split}
\end{align}
In particular, $R_{ij;kl}(K=0)\neq R^\prime_{ij;kl}(K=0)$ if $n_{ik}-n_{jl}\neq 0$.  This indicates that the primed version of Eq.~\eqref{averaged master equation} need not hold given the unprimed version.  However, observe that the diagonal rates $R_{ij;ij}\equiv R_{i\to j}$, along with the diagonal density matrix elements $\rho_{ii}(t)$, are invariant under the gauge transformation \eqref{transformed states and quasienergies} [c.f.~Eq.~\eqref{density matrix transformation}].  Therefore, the rate equation \eqref{full rate equation} that governs the steady-state populations is gauge invariant.

One important consequence of the gauge invariance of Eq.~\eqref{full rate equation} is that, in addition to exact degeneracies where $\epsilon_{ij}=0$, one must also treat carefully \textit{quasi}-degeneracies, where $\epsilon_{ij}=m\omega$ for some $m\in\mathbb Z$.  Such situations were of primary concern in Ref.~\onlinecite{hone}, which points out that such degeneracies can have profound effects on the steady-state density matrix.  In the language of gauge-invariance adopted here, such quasi-degeneracies are simply indicative of the fact that there exists a gauge in which quasienergies $\epsilon_i$ and $\epsilon_j$ are degenerate.  

Consequently, in such situations, Eq.~\eqref{full rate equation} is no longer valid in the basis of Floquet states.  However, the density matrix nevertheless has a block-diagonal structure, $\rho_{ij}$ being nonzero only when $\epsilon_{ij}=0$ in some gauge.  (Let us, for simplicity, adopt this gauge here and for the remainder of the present discussion.)  To close this section, we show that a rate equation of the form \eqref{full rate equation} is recovered upon transforming to the basis in which $\rho_{ij}$ is diagonal.  The unitary transformation $M$ that diagonalizes the density matrix acts on the
Floquet states as
\begin{subequations}
\begin{align}
|u_i(t)\rangle&= 
\sum_{\tilde{i}}\,M_{i\tilde{i}}\,|\tilde{u}_{\tilde i}(t)\rangle
\;,\\
\rho_{ij}&=
\sum_{\tilde{i},\tilde{j}}\;
M^*_{i\tilde{i}}\,M_{j\tilde{j}}\;
\tilde{\rho}_{\tilde{i}\tilde{j}}
\;,\\
S_{ij}&=
\sum_{\tilde{i},\tilde{j}}\;
M^*_{i\tilde{i}}\,M_{j\tilde{j}}\;
\tilde{S}_{\tilde{i}\tilde{j}}
\;,\\
R_{ij;kl}&=
\sum_{\tilde{i},\tilde{j},\tilde{k},\tilde{l}}\;
M^*_{i\tilde{i}}\,M_{j\tilde{j}}\;
M_{k\tilde{k}}\,M^*_{l\tilde{l}}\;
\tilde{R}_{\tilde{i}\tilde{j};\tilde{k}\tilde{l}}
\;.
\end{align}
\end{subequations}
We now plug these transformed quantities into Eq.~\eqref{averaged master equation},
keeping in mind that its left-hand side can be set to zero given that
$\rho_{ij}$ is zero whenever $\epsilon_{ij}\ne m\omega$. We obtain, for each $i$ and $j$,
\begin{align}
\label{eq:hone_2.20_tilde_lin}
0
&=
\sum_{\ti, \tj}
\left(M^*_{i\ti}\;M_{j\tj}\right)
\sum_{\tk, \tl}
\left[
\tilde{\rho}_{\tl \tj}\;\tilde{R}_{\ti \tk ;\tl \tk}
+
\tilde{\rho}_{\ti \tl}\;\tilde{R}^*_{\tj \tk;\tl \tk}\right.\nonumber\\
&\indent\left.
-
\tilde{\rho}_{\tk \tl}\left(\tilde{R}_{\tl \tj ;\tk \ti}+\tilde{R}^*_{\tk \ti ;\tl \tj}\right)
\right]
\;.
\end{align}
One can show that Eq.~(\ref{eq:hone_2.20_tilde_lin}) holds if and only if
\begin{equation}
\label{eq:hone_2.20_tilde}
0
=
\sum_{\tk, \tl}
\left[
\tilde{\rho}_{\tl \tj}\;\tilde{R}_{\ti \tk ;\tl \tk}
+
\tilde{\rho}_{\ti \tl}\;\tilde{R}^*_{\tj \tk;\tl \tk}
-
\tilde{\rho}_{\tk \tl}\left(\tilde{R}_{\tl \tj ;\tk \ti}+\tilde{R}^*_{\tk \ti ;\tl \tj}\right)
\right]
\;,
\end{equation}
for all $\ti$ and $\tj$.  Next, consider the above equation when $\ti=\tj$, while at the same time
keeping in mind that $\tilde\rho$ is diagonal. We obtain
\begin{equation}
0
=
\sum_{\tk}
\tilde{\rho}_{\ti\ti}
\;
\left(
\tilde{R}_{\ti \tk ;\ti \tk}
+
\tilde{R}^*_{\ti \tk;\ti \tk}
\right)
-
\sum_{\tk}
\tilde{\rho}_{\tk\tk}
\left(\tilde{R}_{\tk \ti ;\tk \ti}+\tilde{R}^*_{\tk \ti ;\tk \ti}\right)
\;,
\end{equation}
or equivalently
\begin{equation}
\label{eq:main}
0
=
\sum_{\tk}
\(
\tilde{p}_{\ti}
\;
\tilde{R}_{\ti\to \tk}
-
\tilde{p}_{\tk}
\;
\tilde{R}_{\tk\to\ti}\)
\;,
\end{equation}
where $\tilde{p}_{\ti}$ is the probability of being in the state labeled by
$\ti$, and
\begin{eqnarray}
\tilde{R}_{\ti\to \tk}
&=&
2\pi\gamma^2
\sum_m \left|S_{\ti \tk}(m)\right|^2\,g(\epsilon_{\tk\ti}-m\omega).
\end{eqnarray}
This discussion has therefore demonstrated that a rate equation of the form \eqref{full rate equation} always determines the steady-state reduced density matrix $\rho_{ij}$ in some basis, namely the basis in which $\rho_{ij}$ is diagonal.  Whether or not this basis is the basis of Floquet states, or some other time-periodic basis, depends on the degeneracy structure of the quasienergy spectrum.  Nevertheless, the subsequent analyses presented here hold equally well in this choice of basis.

\subsection{Conditions for thermal Floquet steady states}\label{thermal steady states}

In this section, we will determine, within the master equation formalism discussed above, conditions under which the driven system relaxes to a thermal distribution with respect to the quasienergies.   In particular, we will show that such a situation occurs provided that $P^\dagger(t)H_{\rm SB}P(t)$, where $P(t)$ is defined via Eq.~\eqref{rotating frame} with $H(t)=H_{\rm S}(t)$, is either time-independent or depends on time in a particularly simple way.  While statements to this effect have been made in various works~\cite{iadecola1,iadecola2,shirai,liu}, we provide here a simple and complementary derivation of the statement from the master equation formalism outlined in Ref.~\cite{hone}, and provide connections to the principle of gauge invariance discussed previously.  Before beginning with the derivation, we observe that, at finite temperature, the function $g(E)$, which contains the influence of the bath on the driven system, satisfies
\begin{align}\label{KMS relation}
g(E)=g(-E)\, e^{-\beta E},
\end{align}
which induces the following generalized detailed balance relation for the rates appearing in Eq.~\eqref{full rate equation} \cite{hone}:
\begin{subequations}\label{generalized detailed balance}
\begin{align}
\frac{R_{i\to j}}{R_{j\to i}} = e^{\beta(\epsilon_i-\epsilon_j)}\, \frac{\sum_mR^m_{i\to j}}{\sum_mR^m_{i\to j}e^{-\beta m\omega}},
\end{align}
where 
\begin{align}\label{R^m definition}
R^m_{i\to j}=2\pi\gamma^2\left|S_{i j}(m)\right|^2\,g(\epsilon_{ji}-m\omega).
\end{align}
\end{subequations}
Note that substituting Eq.~\eqref{generalized detailed balance} into the rate equation \eqref{full rate equation} yields a thermal distribution for the occupations of the Floquet states, i.e. $p_j \propto e^{-\beta\, \epsilon_j}$ if $m=0$.

We can now show that such a thermal distribution emerges if $P^\dagger(t)H_{\rm SB}P(t)$ is time-independent, simply by showing that it implies that only the $m=0$ terms contribute to Eq.~\eqref{generalized detailed balance}.  To do this, we make use of an explicit representation of the operator $P(t)$, which can be derived as follows.  Noting first that the evolution operator $U(t_2,t_1)$ can be written in terms of the Floquet states $\ket{u_j(t)}$ as
\begin{align}
\begin{split}
U(t_2,t_1)&=\sum_j \ket{\psi_j(t_2)}\bra{\psi_j(t_1)}\\
&=\sum_j e^{-\mathrm i\epsilon_j(t_2-t_1)}\ket{u_j(t_2)}\bra{u_j(t_1)},
\end{split}
\end{align}
we compute
\begin{align}
\begin{split}
\mathrm i\partial_t U(t,0)&=H_{\rm S}(t)U(t,0)\\
&=\sum_je^{-\mathrm i\epsilon_jt}\Big[\epsilon_j\, \ket{u_j(t)}\bra{u_j(0)}\\
&\qquad\qquad\qquad +\mathrm i\partial_t\ket{u_j(t)}\bra{u_j(0)}\Big].
\end{split}
\end{align}
Acting from the right with $U^\dagger(t,0)$ on both lines above and using the fact that $\braket{u_i(t)|u_j(t)}=\delta_{ij}$ at any time $t$, we deduce that
\begin{align}
H_{\rm S}(t)=\sum_j\, \Big[\epsilon_j\, \ket{u_j(t)}\bra{u_j(t)}+\mathrm i\partial_t\ket{u_j(t)}\bra{u_j(t)}\Big].
\end{align}
At this point, making the ansatz 
\begin{align}\label{explicit P(t)}
P(t) := \sum_j\ket{u_j(t)}\bra{u_j(0)},
\end{align}
we find that indeed
\begin{subequations}
\begin{align}
H_{\rm F}=P^\dagger(t)H_{\rm S}(t)P(t)-\mathrm i\, P^\dagger(t)\partial_t P(t),
\end{align}
where
\begin{align}
H_{\rm F}:=\sum_j \epsilon_j \ket{u_j(0)}\bra{u_j(0)},
\end{align}
\end{subequations}
as desired.

Using the explicit form of $P(t)$ provided in Eq.~\eqref{explicit P(t)}, the desired result follows directly from the definition of $R^m_{i\to j}$ given in Eq.~\eqref{R^m definition}.  In particular, observe that
\begin{align}
\begin{split}
S_{ij}(m)&=\frac{1}{\tau}\int_0^\tau\mathrm d t\, e^{-\mathrm i m\omega t}\, \bra{u_i(t)}S\ket{u_j(t)}\\
&=\frac{1}{\tau}\int_0^\tau\mathrm d t\, e^{-\mathrm i m\omega t}\, \bra{u_i(0)}P^\dagger(t)SP(t)\ket{u_j(0)}.
\end{split}
\end{align}
If $P^\dagger(t)SP(t)$ [and therefore $P^\dagger(t)H_{\rm SB}P(t)$] is independent of time, then $S_{ij}(m)=R^m_{i\to j}=0$ for all $m\neq 0$.  We have therefore shown that, if the operator $P^\dagger(t)H_{\rm SB}P(t)$ is time-independent, then the steady-state occupations of the Floquet states are given by $p_j\propto e^{-\beta\epsilon_j}$, as they would be for a system with Hamiltonian $H_{\rm F}$ at equilibrium with a finite-temperature reservoir.  The zero-temperature limit of the associated grand-canonical distribution determines an unambiguous ordering of the quasienergies.

\begin{figure}
\centering
\includegraphics[width=.45\textwidth]{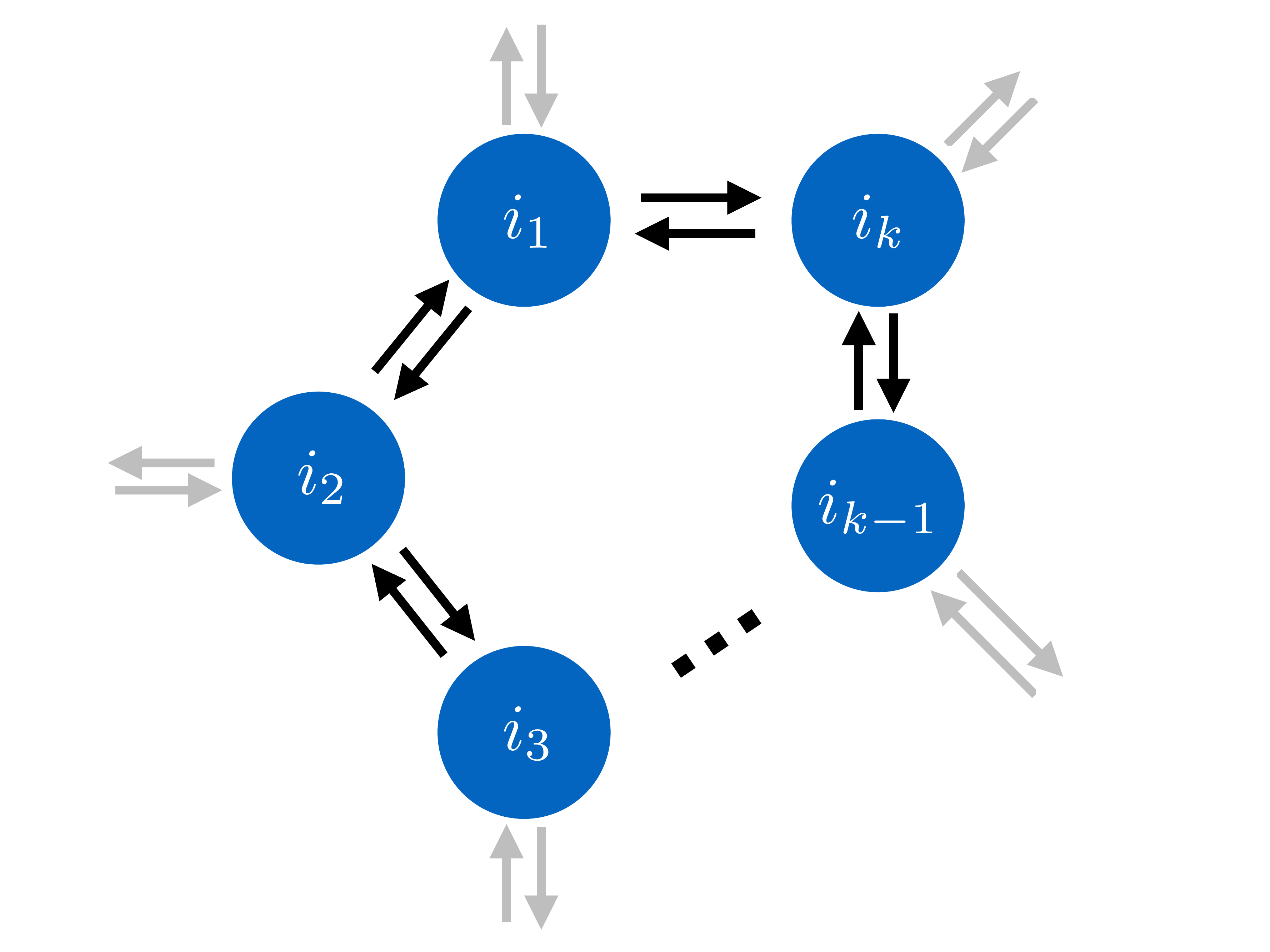}
\caption{(Color online) Closed loop in a graphical representation of the state space defined by the Floquet states and the rates $R_{i\to j}$ connecting them.  Blue circles represent Floquet states, and a directed arrow pointing from state $i$ to state $j$ represents the rate $R_{i\to j}$. If the condition \eqref{gauge condition} is satisfied around all closed loops in the graph, then the extra factors of $m_{ij}\omega$ on the links can be removed by a gauge transformation of the form \eqref{transformed states and quasienergies}, yielding a thermal quasienergy distribution.}\label{bipartite}
\end{figure}

If $P^\dagger(t)SP(t)$ is not time-independent, it is still possible
to reach an effective thermal distribution if detailed balance, namely the relation
\begin{align}\label{detailed balance definition}
\frac{R_{i\to j}}{R_{j\to i}}=\frac{p_j}{p_i},
\end{align}
holds for all $i$ and $j$. For example, let us expand the time-periodic operator
$P^\dagger(t)SP(t)=:\sum_n\,{\cal S}_n\;e^{\mathrm{i} n\omega t}$, and
consider the case where $\bra{u_i(0)}\,{\cal S}_n\,\ket{u_j(0)}$, for fixed $i$ and $j$, is
nonzero for a single mode $n=m_{ij}$. In this case,
\begin{align}
\begin{split}
R_{i\to j}=2\pi\gamma^2\left|S_{i j}(m_{ij})\right|^2\,g(\epsilon_{ji}-m_{ij}\,\omega).
\end{split}
\end{align}
Since $S$ is a Hermitian operator, we have $m_{ij}=-m_{ji}$, and $S_{i
  j}(m_{ij})=S^*_{j i}(m_{ji})$. It follows that
Eq.~\eqref{generalized detailed balance} becomes
\begin{align}\label{detailed balance plus omega}
\frac{R_{i\to j}}{R_{j\to i}} = e^{\beta(\epsilon_i-\epsilon_j+m_{ij}\omega)}.
\end{align}
In this case, one recovers a thermal quasienergy distribution only if it is possible to consistently absorb the extra factor of $m_{ij}\omega$ into a redefinition of the quasienergies $\epsilon_j$. In principle, this can be achieved by means of gauge transformations of the form \eqref{transformed states and quasienergies}, which shift the quasienergies by integer multiples of the driving frequency $\omega$.  In practice, however, it might arise that such a gauge transformation cannot be performed consistently over all Floquet states.  Fortunately, there is a condition, demonstrated below, on the integers $m_{ij}$ that guarantees that such a transformation can be carried out.

The set of nonvanishing transition rates $R_{i\to j}$ defines the directed edges of a graph whose vertices are the Floquet states (see Fig.~\ref{bipartite}).  In order to consistently gauge away the extra factor of $m_{ij}\omega$, it is sufficient to require that for any closed loop
$i_1\to i_2 \to i_3 \to \dots \to i_{k-1}\to i_k\to i_1$ in this graph, one has
\begin{align}
\frac{R_{i_1\to i_2}}{R_{i_2\to i_1}}\frac{R_{i_2\to i_3}}{R_{i_3\to i_2}}\dots \frac{R_{i_{k-1}\to i_k}}{R_{i_k\to i_{k-1}}}\frac{R_{i_{k}\to i_1}}{R_{i_1\to i_{k}}}=1.
\end{align}
[Observe that this condition holds automatically if detailed balance, Eq.~\eqref{detailed balance definition}, holds.]  Using Eq.~\eqref{detailed balance plus omega}, one finds that this condition holds if and only if
\begin{align}\label{gauge condition}
m_{i_1i_2}+m_{i_2i_3}+\dots+m_{i_{k-1}i_k}+m_{i_ki_1}=0,
\end{align}
which is the desired condition on the $m_{ij}$.  The above condition is satisfied if $m_{ij}=m_i-m_j$ for all $i$ and $j$.  [Note, however, that this condition is sufficient but not necessary to satisfy Eq.~\eqref{gauge condition}.]  In this case, one can redefine
the quasienergies via the gauge transformation
\begin{subequations}\label{bipartite construction}
\begin{align}
\epsilon_i\to \epsilon^\prime_i = \epsilon_i+m_i\,\omega
\end{align}
such that Eq.~\eqref{detailed balance plus omega} becomes
\begin{align}
\frac{R_{i\to j}}{R_{j\to i}} &= e^{\beta(\epsilon^\prime_i-\epsilon^\prime_j)},
\end{align}
\end{subequations}
which yields the desired thermal quasienergy distribution $p_i\propto e^{-\beta\epsilon^\prime_i}$.

\section{Zero-temperature nonequilibrium steady states}\label{nonthermal}

In this section, we carry out the program outlined in the introduction, and characterize the zero-temperature occupations of Floquet bands in a generic class of periodically-driven noninteracting systems.  We take the combined Hamiltonian for the system and the reservoir to be given by Eq.~\eqref{model def}, i.e.~
\begin{subequations}
\begin{align}
H(t) = H_{\rm S}(t)+H_{\rm SB}+H_{\rm B}.
\end{align}
The Hamiltonian $H_{\rm S}(t)$ describes the periodically-driven system, which we take for simplicity to be a two-band model of noninteracting fermions. (Generalizing our results to models with more than two bands is straightforward.)  We will focus on the case of monochromatic driving, so that
\begin{align}\label{H_S def}
H_{\rm S}(t)=H_0+\lambda\, H_{\rm D}(t)
\end{align}
where $\lambda$ is the driving amplitude and $H_{\rm D}(t)$ depends on time via linear combinations of $\sin(\omega t)$ and $\cos(\omega t)$.  The quasienergy bands of the driven system, denoted by $\epsilon_j(\bm k)$  ($j=1,2$), are assumed to be gapped (i.e.~nondegenerate for all $\bm k$) and to host nontrivial topological invariants.
The system-bath coupling is again taken to be in the factorized form of Eq.~\eqref{H_SB def}, namely
\begin{align}
H_{\rm SB} = \gamma\, S\, B, 
\end{align}
\end{subequations}
We also take the operator $S$ to conserve both momentum and particle number, so that each $\bm k$ mode is effectively coupled to its own bath, and so hereafter we suppress the momentum index $\bm k$.  Finally, $H_{\rm B}$ describes a bosonic bath, with energy eigenstates $\ket \nu$ and eigenvalues $E_\nu$, which we take to be in equilibrium at zero temperature for all time.  

If the Floquet spectrum is nondegenerate, as we assume, then $\rho_{ij}(t)$ is diagonal at long times, and the steady-state occupation probabilities $p_j := \rho_{jj}(t\to\infty)$, where $p_1+p_2=1$, satisfy the rate equation \eqref{full rate equation}, i.e.
\begin{align}\label{rate equation}
0 = R_{2\to 1}\, p_2 - R_{1\to 2}\, p_1.
\end{align}
We recall that the transition rates are defined as
\begin{subequations}
\begin{align}
R_{i\to j}&:= 2\pi\gamma^2\sum^\infty_{m=-\infty} |S_{ij}(m)|^2\, g(\epsilon_{j}-\epsilon_i-m\omega),\label{transition rates}\\
S_{ij}(m) &:= \frac{1}{\tau}\int_0^\tau\mathrm d t\, e^{-\mathrm i m\omega t}\, \bra{u_i(t)}S\ket{u_j(t)} \label{matrix elements}.
\end{align}
At zero temperature, the weighted bath density of states $g(E)$ defined in Eq.~\eqref{g(E) definition finite T} is given by
\begin{align}\label{g definition}
g(E)=\sum_{\nu} |\langle 0|B|\nu\rangle|^2 \,\delta(E+E_{\nu}).
\end{align}
\end{subequations}
We have set the bath ground-state energy $E_0=0$, so that $E_{\nu>0}$ is strictly positive.  For this reason, $g(E>0)$ vanishes identically at zero temperature, a fact that will be of crucial importance below.  In our analysis, it will be instructive to model $g(E)$ as a power law at low energies compared to a very large cutoff scale \cite{leggett}:
\begin{align}\label{g scaling}
g(E)= g_0\, |E|^\eta\, \theta(-E).
\end{align}
The real exponent $0 < \eta < \infty$ classifies the type of bath; if $\eta = 1$, the bath is referred to as ohmic, while $\eta > 1$ and $\eta < 1$ are referred to as super-ohmic and sub-ohmic, respectively.

It is interesting to note, as pointed out in Sec.~\ref{gauge invariance review}, that the rates entering equation \eqref{rate equation} are invariant under gauge transformations of the form \eqref{gauge}.  While these gauge transformations change the ordering of the quasienergies, they nevertheless do not change the occupations of the Floquet states themselves. However, as we will see later on, an appropriately engineered reservoir is capable of determining an unambiguous ordering of the quasienergies and a choice of an effective ``Floquet ground state."

We now turn to an analysis of the transition rates \eqref{transition rates} that will allow us to determine the extent to which a single Floquet state can be populated at zero temperature, given that the system is coupled to a bath.  For this analysis, it will be convenient to choose a gauge in which the ordering of the quasienergies is determined by the ordering of the energies of the undriven Hamiltonian $H_0$, in such a way that the separation $\Delta := \epsilon_2-\epsilon_1$ is positive.  This gauge can be understood by building up the Floquet states perturbatively in $\lambda$ from the eigenbasis of $H_0$~\cite{sambe}.  To do this, we make use of the Fourier decomposition of the time-periodic Floquet states,
\begin{align}
\ket{u_j(t)}=\sum^\infty_{m=-\infty} e^{\mathrm i m\omega t}\, \ket{u_j^m},
\end{align}
and observe that, to zeroth order in this gauge,
$\ket{u_j^0}$ is nothing but an
eigenstate of $H_0$. (Notice also that $\ket{u_j(n\tau)}=\sum_m
  \ket{u_j^m}$ are the eigenstates of $H_{\rm F}$ underlying the
  Floquet topological insulator.)  The nonzero-$m$ Fourier components
of $\ket{u_j(t)}$ arise due to hybridization of the spectrum via
$H_{\rm D}$, and therefore scale with the driving amplitude $\lambda$
as
\begin{align}
\sqrt{\braket{u^m_j|u_j^m}}\sim (\lambda/\omega)^{|m|},
\end{align}
which follows within perturbation theory to $|m|$-th order in $\lambda$~\cite{footnote_scaling}. Henceforth, we will take $\lambda/\omega$ to be small either on account of a small $\lambda$ or a large $\omega$.  In the Appendix, we examine the case where $\lambda/\omega$ is not small, which is much less favorable for Floquet topological states.  Factoring out the $\lambda/\omega$ scaling from the states $\ket{u^m_i}$, we find that one can rewrite the matrix elements in Eq.~\eqref{matrix elements} in the following form:
\begin{align}\label{S scaling}
\begin{split}
S_{ij}(m) &=\sum^\infty_{k=-\infty}(\lambda/\omega)^{|k|+|k+m|}\bra{u^k_i}S\ket{u^{k+m}_j}\\
&=\(\lambda/\omega\)^{|m|}\, s^m_{ij}\(\lambda/\omega\),
\end{split}
\end{align}
where $s^m_{ij}(\lambda/\omega)$ are regular functions containing the terms in the series that depend weakly on $\lambda/\omega$ or lead to a decay of $S_{ij}(m)$ faster than $(\lambda/\omega)^{|m|}$ as $\lambda/\omega\to 0$.

With all this in mind, we now analyze the quantity
\begin{align}\label{ratio}
\frac{p_2}{p_1} = \frac{R_{1\to 2}}{R_{2\to 1}} = \frac{\sum_m |S_{12}(-m)|^2\, g(\Delta+m\omega)}{\sum_m|S_{12}(+m)|^2\, g(-\Delta+m\omega)},
\end{align}
which is proportional to the density of excitations above the ``lowest" quasienergy state in this gauge ($\epsilon_1$).  If only $m=0$ above contributes, the excitation density vanishes and the lower band is completely filled at long times, as it would be at equilibrium, owing to the fact that the argument of $g(E)$ in the numerator is positive.  While model systems that reach such an effective equilibrium steady state have been studied \cite{breuer,iadecola2,iadecola3,shirai,liu}, it is well-known that these steady states do not occur for generic choices of $H_{\rm SB}$ and $g(E)$.  We ask, instead, whether there are any more general mechanisms or limits that suppress $p_2/p_1$.  The analysis is simplified if we assume that $\lambda/\omega$ is sufficiently small that we can keep only the lowest nontrivial value of $|m|$ in the sums above.  We will focus on the case $\omega > \Delta$, since the opposite case would likely not yield a topological band structure.

For $\omega>\Delta$, then only the terms with $m < 0$ ($m\leq 0$) contribute to the numerator (denominator) of Eq.~\eqref{ratio}.  In this case, we find that suppression of $p_2/p_1$ is possible in the limit $\omega\gg \Delta$, which yields [c.f.~Eq.~\eqref{S scaling}]
\begin{align}
\frac{p_2}{p_1}\approx \[\frac{|s^{-1}_{12}(\lambda/\omega)|^2}{|s^1_{12}(\lambda/\omega)|^2}+\frac{|s^0_{12}(\lambda/\omega)|^2}{|s^1_{12}(\lambda/\omega)|^2}\, \(\frac{\omega}{\lambda}\)^2 \,\(\frac{\Delta}{\omega}\)^\eta\]^{-1}.
\end{align}
In addition to the exponent $\eta$, the behavior of $p_2/p_1$ as $\omega\to\infty$ depends on the scaling of the quasienergy separation $\Delta$ with $\lambda$ and $\omega$.  We assume the scaling $\Delta \sim \lambda\, (\lambda/\omega)^\alpha$, for $\alpha\geq 0$; for example, the case $\alpha=1$ corresponds to the size of the direct gap predicted in graphene coupled to a circularly polarized electric field~\cite{oka,kitagawa}.  We additionally allow the driving amplitude to scale with the frequency, $\lambda \sim \omega^\beta$, as it may in some physical driven systems~\cite{bukov,aidelsburger}.  Using these scaling forms, we find that as $\omega\to\infty$ the excitation density
\begin{subequations}\label{power law}
\begin{align}
\frac{p_2}{p_1}\sim \omega^{-(1-\beta)\[2-\eta(\alpha+1)\]} \to 0,
\end{align}
so long as the product
\begin{align}
(1-\beta)\[2-\eta(\alpha+1)\]>0.
\end{align}
Noting that the simultaneous requirement of small $\lambda/\omega$ and large $\omega$ restricts $\beta\leq1$, this criterion reduces to
\begin{align}\label{criterion}
\eta < \frac{2}{\alpha+1}.
\end{align}
\end{subequations}
If this condition is not satisfied, then $p_2/p_1$ approaches a nonuniversal constant in the high-frequency limit, and no single Floquet state is fully populated in the steady state.  It is important to note that $\alpha$, and therefore the excitation density, is $\bm k$-dependent, since the scaling of $\Delta$ with $\lambda/\omega$ varies in momentum space.  In particular, for $\bm k$ far away from the value at which the minimal quasienergy separation occurs, $\Delta$ becomes independent of $\lambda$ and $\omega$.  The $\omega$-scaling for this case is obtained by setting $\alpha=0$ in Eqs.~\eqref{power law}.

For a given $\alpha$ and $\beta$ (fixed by the physical realization of the system), the above result suggests that the low-energy behavior of the function $g(E)$ essentially determines whether or not a single Floquet band is occupied in the limit $\omega \gg \Delta$.  For example, in the case of graphene in a circularly-polarized electric field ($\alpha=1$, $\beta=0$), there is a critical value $\eta_{\rm c} = 1$ (i.e., ohmic dissipation) that separates the power-law decay of $p_2/p_1$ from the aforementioned nonuniversal behavior.  Therefore, an ohmic bath already violates Eq.~\eqref{criterion} for graphene in a circularly-polarized electric field, and the population of the bands near the $K$-point is not controllable by increasing $\omega$ if the bath is ohmic.

In cases where the excitation density decays as a power law at large frequencies, one must still take care to determine whether the resulting steady state has the desired characteristics of the topological Floquet bands.  As $\omega$ increases, the Floquet effective Hamiltonian $H_{\rm F}$ may approach $H_0$ as $1/\omega$ or faster.  If the power-law decay of $p_2/p_1$ is faster than this approach, then the suppression of excitations can still occur in a regime where the Floquet bands are topological.  The situation can be improved by allowing a scaling of the amplitude $\lambda\sim\omega^\beta$ for $\beta\neq 0$~\cite{bukov}, but the value of the exponent $\beta$ must be balanced against an appropriate value of $\eta$ [c.f.~Eq.~\eqref{criterion}] in order for the desired suppression to take place.  Furthermore, it is important to keep in mind that, depending on the exponents $\alpha,\beta,$ and $\eta$, the power law decay of $p_2/p_1$ can be very slow, so that a finite density of excitations remains even at high frequencies compared to all system energy scales.  In order for excitations to be completely suppressed, one must have
\begin{subequations}
\begin{align}
T_{\rm eff} < \Delta,
\end{align}
where the frequency- and momentum-dependent effective temperature $T_{\rm eff}$ is defined as
\begin{align}
T_{\rm eff} = \frac{\Delta}{\ln (p_1/p_2)}.
\end{align}
\end{subequations}
Note that this effective temperature arises even though the bath itself is at zero temperature, and can therefore be understood as a signature of heating effects due to the driving.

\section{The role of bath engineering}
The sensitive dependence on the exponent $\eta$ of the large-frequency scaling of the excitation density already demonstrates the crucial role that the bath plays in stabilizing Floquet topological states in open systems.  We close by commenting on two additional ways in which the bath and its coupling to the system can be engineered in order to further favor the suppression of $p_2/p_1$.  

\begin{figure}
\centering
\includegraphics[width=.35\textwidth,page=1]{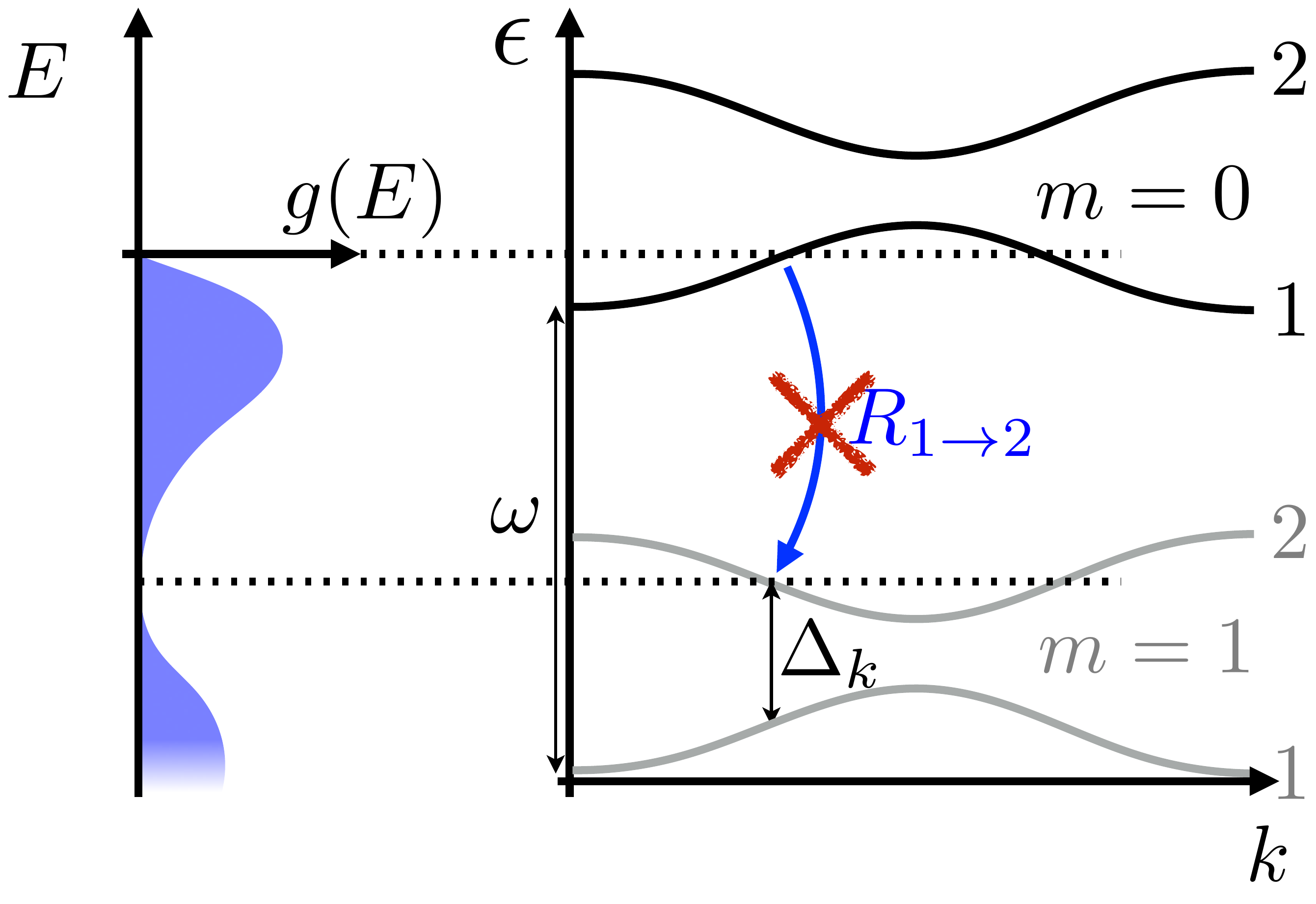}
\caption{(Color online) Suppressing unwanted transitions between Floquet states by design of the function $g(E)$.  Depleting the density of states in the neighborhood of such a transition ensures that there is no corresponding transition in the bath to compensate, essentially forbidding it.}\label{fig:dip}
\end{figure}

One way to suppress excitations is to engineer the spectrum of the bath itself so that the function $g(E)$ appearing in Eq.~\eqref{transition rates} does not have a simple power-law form as in Eq.~\eqref{g scaling}, but instead drops to zero in a neighborhood of $E=\Delta_0-\omega$, where $\Delta_0:=\Delta(\bm k_0)$ is some reference value of the quasienergy separation, as in Fig.~\ref{fig:dip}.  (Such a scenario could be envisioned if, for example, the bath consists of quantized electromagnetic radiation in a cavity, whose size could be tuned to achieve the desired effect.)  If the width of the dip in $g(E)$ is on the order of the width of the upper band, then excitations can be suppressed even if $\omega$ is not much larger than $\Delta_0$.   Indeed, if $\omega > \Delta_0$ and $\lambda/\omega$ is sufficiently small that only the first few terms in the sums over $m$ in Eq.~\eqref{ratio} are kept, then the suppression of $g(E)$ near this value eliminates the terms with $m\neq0$ up to order $(\lambda/\omega)^4$.  However, it is important to point out that the excitation density in this case still exhibits at best power-law decay at large frequencies, with appropriate modifications to Eqs.~\eqref{power law} arising from keeping terms other than $|m|=0,1$ in Eq.~\eqref{ratio}.  To completely eradicate excitations to all orders in $\lambda/\omega$, one must engineer dips at energies $E=\Delta_0-m\omega$ for all $m>0$.   

Of course, by placing the dip at $E=-\Delta_0-\omega$, one can also use this mechanism to populate what we have referred to as the upper band in this choice of gauge.  While this scenario looks like a population inversion, one can of course perform an appropriate transformation of the form \eqref{gauge} to reorder the Floquet bands in such a way that the populated band is the lowest.  This example indicates that, in certain cases, the reservoir can ``choose" a preferred gauge in which the system appears to be (nearly) at equilibrium.

The system-bath coupling is another quantity that could potentially be manipulated in order to suppress excitations.  Indeed, certain system-bath couplings are known to yield relaxation to steady states that feature filled Floquet bands.  For example, if $H_{\rm SB}$ is chosen in such a way that $P^\dagger(t)H_{\rm SB} P(t)$, where $P(t)$ satisfies Eq.~\eqref{rotating frame} with $H(t)=H_{\rm S}(t)$, is time-independent, then the system described by the total Hamiltonian $H(t)$ defined in Eq.~\eqref{model def} reaches an effective thermal equilibrium with respect to the eigenvalues of $H_{\rm F}$ (see Sec.~\ref{thermal steady states} and Refs.~\cite{iadecola1,iadecola2,shirai,liu}).  Given sufficient control over the system-bath coupling, one could attempt to engineer such a situation, at least to some order in $\lambda/\omega$, by designing an $H_{\rm SB}$ that, say, commutes with the lowest nontrivial Fourier harmonic of $P(t)$, or even by engineering an appropriate time dependence in $H_{\rm SB}$ to cancel the time-dependence in $P(t)$ to some order.

\section{Summary and Conclusion}

We have argued in this work that the possibility of stabilizing a Floquet topological state with a low density of excitations is heavily constrained by the coupling to a thermal reservoir.  Using scaling arguments, we demonstrated that, even in the limit of weak driving and/or high frequenscy, the bath density of states has tremendous influence on whether or not excitations are suppressed as $\omega\to\infty$.  We also suggested ways of designing the bath and its coupling to the system in order to suppress excitations.

Our results suggest that it is at best difficult, and at worst impossible, to engineer a periodically-driven quantum system whose steady state resembles the zero-temperature ground state of some target topological phase.  However, even out of equilibrium, there is reason to believe that nontrivial features, such as topological indices~\cite{d'alessio}, edge states~\cite{moore}, and (approximately) quantized transport~\cite{dehghani2,torres}, survive in both isolated and open systems.  Indeed, there is already experimental evidence to this effect in cold atomic gases~\cite{aidelsburger,jotzu}.  We emphasize, however, that it is precisely in the deviations from the resemblance to equilibrium systems where the newest physics lies.  For example, interacting versions of these models~\cite{grushin} could be used as platforms to probe fractionalized excitations out of equilibrium.

\section*{Acknowledgments}
We thank Camille Aron and Garry Goldstein for inspiring discussions.  T.I. was supported by the National Science Foundation Graduate Research Fellowship Program under Grant No.~DGE-1247312. T.N. was supported by DARPA
SPAWARSYSCEN Pacific N66001-11-1-4110, and C.C. was supported by DOE Grant DEF-06ER46316.\\

{\it Note added ---} During preparation of this manuscript, we became aware of Ref.~\onlinecite{seetharam}, which also discusses the possibility of stabilizing Floquet topological states with couplings to particular appropriately-engineered baths.

\appendix*

\section{Limit of strong driving}
\setcounter{equation}{0}
\renewcommand{\theequation}{A\arabic{equation}}

In the case of strong driving ($\lambda/\omega\gg 1$), there are many values of $m$ for which $S_{ij}(m)$ can be non-negligible.  Indeed, as $\lambda/\omega\to\infty$, the Floquet states can become chaotic, so that the $S_{ij}(m)$ may be regarded as essentially random variables, whose magnitudes need not decay quickly as $|m|$ becomes large.  For this reason, the sums in the numerator and denominator of Eq.~\eqref{ratio} generically diverge in the limit $\lambda/\omega \to \infty$, and the ratio of transition rates is indeterminate.  One can, however, identify constraints on the amplitudes $|S_{12}(m)|^2$ such that the ratio converges to a definite finite value.  In particular, if
\begin{align}
|S_{12}(\pm |m|)|^2 < (\text{const.})\times \frac{1}{|m|^{1+\eta+\delta}}
\end{align}
as $|m|\to\infty$ for any positive real number $\delta$, then both series are bounded from above by a convergent series, and therefore the ratio has a definite value.  This is true even for infinitesimally small $\delta\to 0_+$.

Even if the sums in the numerator and denominator are divergent, the ratio \eqref{ratio} can approach a finite value for system-bath coupling operators $S$ such that $|S_{ij}(m)|^2=|S_{ij}(-m)|^2$, due to a symmetry.  To see this, let us drop the $m=0$ term in the denominator and rewrite Eq.~\eqref{ratio} for $\omega > \Delta$ as
\begin{align}
\frac{p_2}{p_1}&\approx \frac{\sum_{m<0} |S_{12}(|m|)|^2\, (|m|\omega-\Delta)^\eta}{\sum_{m<0}|S_{12}(|m|)|^2\, (|m|\omega+\Delta)^\eta}.
\end{align}
For large $|m|$, the summands in the numerator and denominator become identical.  Therefore, if $|S_{12}(|m|)|^2$ is finite for sufficiently large $|m|$, the ratio of the two sums approaches 1 from below as $|m|$ grows.  When this occurs, the system approaches an infinite effective temperature --- all Floquet states are occupied with equal probabilities, despite the fact that the bath is held at zero temperature.  If there exists some $|m_\text{max}|$ such that $|S_{12}(|m_{\rm max}|)|^2=0$, then the ratio takes on a finite value that is bounded from above by unity.

\bibliographystyle{apsrev}

\bibliography{refs_gapped_Floquet}

\begin{thebibliography}{36}
\expandafter\ifx\csname natexlab\endcsname\relax\def\natexlab#1{#1}\fi
\expandafter\ifx\csname bibnamefont\endcsname\relax
  \def\bibnamefont#1{#1}\fi
\expandafter\ifx\csname bibfnamefont\endcsname\relax
  \def\bibfnamefont#1{#1}\fi
\expandafter\ifx\csname citenamefont\endcsname\relax
  \def\citenamefont#1{#1}\fi
\expandafter\ifx\csname url\endcsname\relax
  \def\url#1{\texttt{#1}}\fi
\expandafter\ifx\csname urlprefix\endcsname\relax\def\urlprefix{URL }\fi
\providecommand{\bibinfo}[2]{#2}
\providecommand{\eprint}[2][]{\url{#2}}

\bibitem[{\citenamefont{Oka and Aoki}(2009)}]{oka}
\bibinfo{author}{\bibfnamefont{T.}~\bibnamefont{Oka}} \bibnamefont{and}
  \bibinfo{author}{\bibfnamefont{H.}~\bibnamefont{Aoki}},
  \bibinfo{journal}{Phys. Rev. B} \textbf{\bibinfo{volume}{79}},
  \bibinfo{pages}{081406} (\bibinfo{year}{2009}).

\bibitem[{\citenamefont{Lindner et~al.}(2011)\citenamefont{Lindner, Refael, and
  Galitski}}]{lindner}
\bibinfo{author}{\bibfnamefont{N.~H.} \bibnamefont{Lindner}},
  \bibinfo{author}{\bibfnamefont{G.}~\bibnamefont{Refael}}, \bibnamefont{and}
  \bibinfo{author}{\bibfnamefont{V.}~\bibnamefont{Galitski}},
  \bibinfo{journal}{Nature Phys.} \textbf{\bibinfo{volume}{7}},
  \bibinfo{pages}{490} (\bibinfo{year}{2011}).

\bibitem[{\citenamefont{Kitagawa et~al.}(2011)\citenamefont{Kitagawa, Oka,
  Brataas, Fu, and Demler}}]{kitagawa}
\bibinfo{author}{\bibfnamefont{T.}~\bibnamefont{Kitagawa}},
  \bibinfo{author}{\bibfnamefont{T.}~\bibnamefont{Oka}},
  \bibinfo{author}{\bibfnamefont{A.}~\bibnamefont{Brataas}},
  \bibinfo{author}{\bibfnamefont{L.}~\bibnamefont{Fu}}, \bibnamefont{and}
  \bibinfo{author}{\bibfnamefont{E.}~\bibnamefont{Demler}},
  \bibinfo{journal}{Phys. Rev. B} \textbf{\bibinfo{volume}{84}},
  \bibinfo{pages}{235108} (\bibinfo{year}{2011}).

\bibitem[{\citenamefont{Gu et~al.}(2011)\citenamefont{Gu, Fertig, Arovas, and
  Auerbach}}]{fertig}
\bibinfo{author}{\bibfnamefont{Z.}~\bibnamefont{Gu}},
  \bibinfo{author}{\bibfnamefont{H.~A.} \bibnamefont{Fertig}},
  \bibinfo{author}{\bibfnamefont{D.~P.} \bibnamefont{Arovas}},
  \bibnamefont{and} \bibinfo{author}{\bibfnamefont{A.}~\bibnamefont{Auerbach}},
  \bibinfo{journal}{Phys. Rev. Lett.} \textbf{\bibinfo{volume}{107}},
  \bibinfo{pages}{216601} (\bibinfo{year}{2011}).

\bibitem[{\citenamefont{Rudner et~al.}(2013)\citenamefont{Rudner, Lindner,
  Berg, and Levin}}]{rudner}
\bibinfo{author}{\bibfnamefont{M.~S.} \bibnamefont{Rudner}},
  \bibinfo{author}{\bibfnamefont{N.~H.} \bibnamefont{Lindner}},
  \bibinfo{author}{\bibfnamefont{E.}~\bibnamefont{Berg}}, \bibnamefont{and}
  \bibinfo{author}{\bibfnamefont{M.}~\bibnamefont{Levin}},
  \bibinfo{journal}{Phys. Rev. X} \textbf{\bibinfo{volume}{3}},
  \bibinfo{pages}{031005} (\bibinfo{year}{2013}).

\bibitem[{\citenamefont{Grushin et~al.}(2014)\citenamefont{Grushin,
  G\'omez-Le\'on, and Neupert}}]{grushin}
\bibinfo{author}{\bibfnamefont{A.~G.} \bibnamefont{Grushin}},
  \bibinfo{author}{\bibfnamefont{A.}~\bibnamefont{G\'omez-Le\'on}},
  \bibnamefont{and} \bibinfo{author}{\bibfnamefont{T.}~\bibnamefont{Neupert}},
  \bibinfo{journal}{Phys. Rev. Lett.} \textbf{\bibinfo{volume}{112}},
  \bibinfo{pages}{156801} (\bibinfo{year}{2014}).

\bibitem[{\citenamefont{Kundu et~al.}(2014)\citenamefont{Kundu, Fertig, and
  Seradjeh}}]{kundu}
\bibinfo{author}{\bibfnamefont{A.}~\bibnamefont{Kundu}},
  \bibinfo{author}{\bibfnamefont{H.~A.} \bibnamefont{Fertig}},
  \bibnamefont{and} \bibinfo{author}{\bibfnamefont{B.}~\bibnamefont{Seradjeh}},
  \bibinfo{journal}{Phys. Rev. Lett.} \textbf{\bibinfo{volume}{113}},
  \bibinfo{pages}{236803} (\bibinfo{year}{2014}).

\bibitem[{\citenamefont{Foa~Torres et~al.}(2014)\citenamefont{Foa~Torres,
  Perez-Piskunow, Balseiro, and Usaj}}]{torres}
\bibinfo{author}{\bibfnamefont{L.~E.~F.} \bibnamefont{Foa~Torres}},
  \bibinfo{author}{\bibfnamefont{P.~M.} \bibnamefont{Perez-Piskunow}},
  \bibinfo{author}{\bibfnamefont{C.~A.} \bibnamefont{Balseiro}},
  \bibnamefont{and} \bibinfo{author}{\bibfnamefont{G.}~\bibnamefont{Usaj}},
  \bibinfo{journal}{Phys. Rev. Lett.} \textbf{\bibinfo{volume}{113}},
  \bibinfo{pages}{266801} (\bibinfo{year}{2014}).

\bibitem[{moo()}]{moore}
\bibinfo{howpublished}{J.~P. Dahlhaus, B.~M. Fregoso, and J.~E. Moore,
  arXiv:1408.6811 (unpublished).}

\bibitem[{\citenamefont{Shirley}(1965)}]{shirley}
\bibinfo{author}{\bibfnamefont{J.~H.} \bibnamefont{Shirley}},
  \bibinfo{journal}{Phys. Rev.} \textbf{\bibinfo{volume}{138}},
  \bibinfo{pages}{B979} (\bibinfo{year}{1965}).

\bibitem[{\citenamefont{Sambe}(1973)}]{sambe}
\bibinfo{author}{\bibfnamefont{H.}~\bibnamefont{Sambe}},
  \bibinfo{journal}{Phys. Rev. A} \textbf{\bibinfo{volume}{7}},
  \bibinfo{pages}{2203} (\bibinfo{year}{1973}).

\bibitem[{\citenamefont{Rahav et~al.}(2003)\citenamefont{Rahav, Gilary, and
  Fishman}}]{rahav}
\bibinfo{author}{\bibfnamefont{S.}~\bibnamefont{Rahav}},
  \bibinfo{author}{\bibfnamefont{I.}~\bibnamefont{Gilary}}, \bibnamefont{and}
  \bibinfo{author}{\bibfnamefont{S.}~\bibnamefont{Fishman}},
  \bibinfo{journal}{Phys. Rev. A} \textbf{\bibinfo{volume}{68}},
  \bibinfo{pages}{013820} (\bibinfo{year}{2003}).

\bibitem[{\citenamefont{Goldman and Dalibard}(2014)}]{goldman}
\bibinfo{author}{\bibfnamefont{N.}~\bibnamefont{Goldman}} \bibnamefont{and}
  \bibinfo{author}{\bibfnamefont{J.}~\bibnamefont{Dalibard}},
  \bibinfo{journal}{Phys. Rev. X} \textbf{\bibinfo{volume}{4}},
  \bibinfo{pages}{031027} (\bibinfo{year}{2014}).

\bibitem[{buk()}]{bukov}
\bibinfo{howpublished}{M. Bukov, L. D'Alessio, and A. Polkovnikov,
  arXiv:1407.4803 (unpublished).}

\bibitem[{\citenamefont{Hasan and Kane}(2010)}]{hasan}
\bibinfo{author}{\bibfnamefont{M.~Z.} \bibnamefont{Hasan}} \bibnamefont{and}
  \bibinfo{author}{\bibfnamefont{C.~L.} \bibnamefont{Kane}},
  \bibinfo{journal}{Rev. Mod. Phys.} \textbf{\bibinfo{volume}{82}},
  \bibinfo{pages}{3045} (\bibinfo{year}{2010}).

\bibitem[{\citenamefont{Kohn}(2001)}]{kohn}
\bibinfo{author}{\bibfnamefont{W.}~\bibnamefont{Kohn}}, \bibinfo{journal}{J.
  Stat. Phys.} \textbf{\bibinfo{volume}{103}}, \bibinfo{pages}{417}
  (\bibinfo{year}{2001}).

\bibitem[{\citenamefont{Dehghani et~al.}(2014)\citenamefont{Dehghani, Oka, and
  Mitra}}]{dehghani1}
\bibinfo{author}{\bibfnamefont{H.}~\bibnamefont{Dehghani}},
  \bibinfo{author}{\bibfnamefont{T.}~\bibnamefont{Oka}}, \bibnamefont{and}
  \bibinfo{author}{\bibfnamefont{A.}~\bibnamefont{Mitra}},
  \bibinfo{journal}{Phys. Rev. B} \textbf{\bibinfo{volume}{90}},
  \bibinfo{pages}{195429} (\bibinfo{year}{2014}).

\bibitem[{deh()}]{dehghani2}
\bibinfo{howpublished}{H. Dehghani, T. Oka, and A. Mitra, arXiv:1412.8469
  (unpublished).}

\bibitem[{gri()}]{grifoni}
\bibinfo{howpublished}{M. Grifoni and P. H\"anggi, Phys. Rep. \textbf{304}, 229
  (1998).}

\bibitem[{\citenamefont{Kohler et~al.}(1997)\citenamefont{Kohler, Dittrich, and
  H\"anggi}}]{kohler}
\bibinfo{author}{\bibfnamefont{S.}~\bibnamefont{Kohler}},
  \bibinfo{author}{\bibfnamefont{T.}~\bibnamefont{Dittrich}}, \bibnamefont{and}
  \bibinfo{author}{\bibfnamefont{P.}~\bibnamefont{H\"anggi}},
  \bibinfo{journal}{Phys. Rev. E} \textbf{\bibinfo{volume}{55}},
  \bibinfo{pages}{300} (\bibinfo{year}{1997}).

\bibitem[{\citenamefont{Breuer et~al.}(2000)\citenamefont{Breuer, Huber, and
  Petruccione}}]{breuer}
\bibinfo{author}{\bibfnamefont{H.-P.} \bibnamefont{Breuer}},
  \bibinfo{author}{\bibfnamefont{W.}~\bibnamefont{Huber}}, \bibnamefont{and}
  \bibinfo{author}{\bibfnamefont{F.}~\bibnamefont{Petruccione}},
  \bibinfo{journal}{Phys. Rev. E} \textbf{\bibinfo{volume}{61}},
  \bibinfo{pages}{4883} (\bibinfo{year}{2000}).

\bibitem[{\citenamefont{Kohler et~al.}(2005)\citenamefont{Kohler, Lehmann, and
  H\"anggi}}]{kohler_review}
\bibinfo{author}{\bibfnamefont{S.}~\bibnamefont{Kohler}},
  \bibinfo{author}{\bibfnamefont{J.}~\bibnamefont{Lehmann}}, \bibnamefont{and}
  \bibinfo{author}{\bibfnamefont{P.}~\bibnamefont{H\"anggi}},
  \bibinfo{journal}{Phys. Rep.} \textbf{\bibinfo{volume}{406}},
  \bibinfo{pages}{379 } (\bibinfo{year}{2005}).

\bibitem[{\citenamefont{Hone et~al.}(2009)\citenamefont{Hone, Ketzmerick, and
  Kohn}}]{hone}
\bibinfo{author}{\bibfnamefont{D.~W.} \bibnamefont{Hone}},
  \bibinfo{author}{\bibfnamefont{R.}~\bibnamefont{Ketzmerick}},
  \bibnamefont{and} \bibinfo{author}{\bibfnamefont{W.}~\bibnamefont{Kohn}},
  \bibinfo{journal}{Phys. Rev. E} \textbf{\bibinfo{volume}{79}},
  \bibinfo{pages}{051129} (\bibinfo{year}{2009}).

\bibitem[{gra()}]{grand_floquet}
\bibinfo{howpublished}{T. Iadecola and C. Chamon, arXiv:1412.5599
  (unpublished).}

\bibitem[{see()}]{seetharam}
\bibinfo{howpublished}{K.~I. Seetharam, C.-E. Bardyn, N.~H. Lindner, M.~S.
  Rudner, and G. Refael, arXiv:1502.02664 (unpublished).}

\bibitem[{foo({\natexlab{a}})}]{footnote_sys-bath}
\bibinfo{howpublished}{The alternative choice $H_{\rm
  SB}=\sum_{j,\nu}\(\gamma_{j\nu}\, S_j B_\nu+\text{H.c.}\)$, which is
  sufficiently general to include couplings to phonons and fermionic leads,
  results only in cosmetic changes to the resulting kinetic equations; we
  therefore opt instead for the simpler form in Eq.~\eqref{H_SB def}.}

\bibitem[{\citenamefont{Leggett et~al.}(1987)\citenamefont{Leggett,
  Chakravarty, Dorsey, Fisher, Garg, and Zwerger}}]{leggett}
\bibinfo{author}{\bibfnamefont{A.~J.} \bibnamefont{Leggett}},
  \bibinfo{author}{\bibfnamefont{S.}~\bibnamefont{Chakravarty}},
  \bibinfo{author}{\bibfnamefont{A.~T.} \bibnamefont{Dorsey}},
  \bibinfo{author}{\bibfnamefont{M.~P.~A.} \bibnamefont{Fisher}},
  \bibinfo{author}{\bibfnamefont{A.}~\bibnamefont{Garg}}, \bibnamefont{and}
  \bibinfo{author}{\bibfnamefont{W.}~\bibnamefont{Zwerger}},
  \bibinfo{journal}{Rev. Mod. Phys.} \textbf{\bibinfo{volume}{59}},
  \bibinfo{pages}{1} (\bibinfo{year}{1987}).

\bibitem[{\citenamefont{Iadecola
  et~al.}(2013{\natexlab{a}})\citenamefont{Iadecola, Campbell, Chamon, Hou,
  Jackiw, Pi, and Kusminskiy}}]{iadecola1}
\bibinfo{author}{\bibfnamefont{T.}~\bibnamefont{Iadecola}},
  \bibinfo{author}{\bibfnamefont{D.}~\bibnamefont{Campbell}},
  \bibinfo{author}{\bibfnamefont{C.}~\bibnamefont{Chamon}},
  \bibinfo{author}{\bibfnamefont{C.-Y.} \bibnamefont{Hou}},
  \bibinfo{author}{\bibfnamefont{R.}~\bibnamefont{Jackiw}},
  \bibinfo{author}{\bibfnamefont{S.-Y.} \bibnamefont{Pi}}, \bibnamefont{and}
  \bibinfo{author}{\bibfnamefont{S.~V.} \bibnamefont{Kusminskiy}},
  \bibinfo{journal}{Phys. Rev. Lett.} \textbf{\bibinfo{volume}{110}},
  \bibinfo{pages}{176603} (\bibinfo{year}{2013}{\natexlab{a}}).

\bibitem[{\citenamefont{Iadecola
  et~al.}(2013{\natexlab{b}})\citenamefont{Iadecola, Chamon, Jackiw, and
  Pi}}]{iadecola2}
\bibinfo{author}{\bibfnamefont{T.}~\bibnamefont{Iadecola}},
  \bibinfo{author}{\bibfnamefont{C.}~\bibnamefont{Chamon}},
  \bibinfo{author}{\bibfnamefont{R.}~\bibnamefont{Jackiw}}, \bibnamefont{and}
  \bibinfo{author}{\bibfnamefont{S.-Y.} \bibnamefont{Pi}},
  \bibinfo{journal}{Phys. Rev. B} \textbf{\bibinfo{volume}{88}},
  \bibinfo{pages}{104302} (\bibinfo{year}{2013}{\natexlab{b}}).

\bibitem[{\citenamefont{Shirai et~al.}(2015)\citenamefont{Shirai, Mori, and
  Miyashita}}]{shirai}
\bibinfo{author}{\bibfnamefont{T.}~\bibnamefont{Shirai}},
  \bibinfo{author}{\bibfnamefont{T.}~\bibnamefont{Mori}}, \bibnamefont{and}
  \bibinfo{author}{\bibfnamefont{S.}~\bibnamefont{Miyashita}},
  \bibinfo{journal}{Phys. Rev. E} \textbf{\bibinfo{volume}{91}},
  \bibinfo{pages}{030101} (\bibinfo{year}{2015}).

\bibitem[{\citenamefont{Liu}(2015)}]{liu}
\bibinfo{author}{\bibfnamefont{D.~E.} \bibnamefont{Liu}},
  \bibinfo{journal}{Phys. Rev. B} \textbf{\bibinfo{volume}{91}},
  \bibinfo{pages}{144301} (\bibinfo{year}{2015}).

\bibitem[{foo({\natexlab{b}})}]{footnote_scaling}
\bibinfo{howpublished}{Note that one cannot rule out the possibility that
  $\sqrt{\braket{u^m_j|u_j^m}}$ scales with an additional nonuniversal function
  of $\omega$ that approaches unity as $\omega\to\infty$.}

\bibitem[{\citenamefont{Iadecola et~al.}(2014)\citenamefont{Iadecola, Neupert,
  and Chamon}}]{iadecola3}
\bibinfo{author}{\bibfnamefont{T.}~\bibnamefont{Iadecola}},
  \bibinfo{author}{\bibfnamefont{T.}~\bibnamefont{Neupert}}, \bibnamefont{and}
  \bibinfo{author}{\bibfnamefont{C.}~\bibnamefont{Chamon}},
  \bibinfo{journal}{Phys. Rev. B} \textbf{\bibinfo{volume}{89}},
  \bibinfo{pages}{115425} (\bibinfo{year}{2014}).

\bibitem[{\citenamefont{Aidelsburger et~al.}(2013)\citenamefont{Aidelsburger,
  Atala, Lohse, Barreiro, Paredes, and Bloch}}]{aidelsburger}
\bibinfo{author}{\bibfnamefont{M.}~\bibnamefont{Aidelsburger}},
  \bibinfo{author}{\bibfnamefont{M.}~\bibnamefont{Atala}},
  \bibinfo{author}{\bibfnamefont{M.}~\bibnamefont{Lohse}},
  \bibinfo{author}{\bibfnamefont{J.}~\bibnamefont{Barreiro}},
  \bibinfo{author}{\bibfnamefont{B.}~\bibnamefont{Paredes}}, \bibnamefont{and}
  \bibinfo{author}{\bibfnamefont{I.}~\bibnamefont{Bloch}},
  \bibinfo{journal}{Phys. Rev. Lett.} \textbf{\bibinfo{volume}{111}},
  \bibinfo{pages}{185301} (\bibinfo{year}{2013}).

\bibitem[{d'a()}]{d'alessio}
\bibinfo{howpublished}{L. D'Alessio and M. Rigol, arXiv:1409.6319
  (unpublished).}

\bibitem[{\citenamefont{Jotzu et~al.}(2014)\citenamefont{Jotzu, Messer,
  Desbuquois, Lebrat, Uehlinger, Greif, and Esslinger}}]{jotzu}
\bibinfo{author}{\bibfnamefont{G.}~\bibnamefont{Jotzu}},
  \bibinfo{author}{\bibfnamefont{M.}~\bibnamefont{Messer}},
  \bibinfo{author}{\bibfnamefont{R.}~\bibnamefont{Desbuquois}},
  \bibinfo{author}{\bibfnamefont{M.}~\bibnamefont{Lebrat}},
  \bibinfo{author}{\bibfnamefont{T.}~\bibnamefont{Uehlinger}},
  \bibinfo{author}{\bibfnamefont{D.}~\bibnamefont{Greif}}, \bibnamefont{and}
  \bibinfo{author}{\bibfnamefont{T.}~\bibnamefont{Esslinger}},
  \bibinfo{journal}{Nature} \textbf{\bibinfo{volume}{515}},
  \bibinfo{pages}{237} (\bibinfo{year}{2014}).

\end{thebibliography}

\end{document}